# Simulation of the electron coherent radiation process in a crystalline undulator

**N.F. Shul'ga[a,b,][*] and V.I.Truten'[a]**

[a] *Akhiezer Institute for Theoretical Physics of National Science Center "Kharkov Institute of Physics and Technology", Akademicheskaya Str.,1, 61108 Kharkov, Ukraine*
[b] *V.N.Karazin Kharkov National University, Svobody Sq. 4, 61022 Kharkov, Ukraine*
   *E-mail*: shulga@kipt.kharkov.ua

ABSTRACT: When a high-energy electrons move in a crystal along one of the crystallographic planes the phenomenon of planar channeling is possible, in which the particles move in channels formed by the continuous potential of crystalline planes. Such particle motion in a crystal is conserved for most particles even for weak bending of crystalline planes of atoms. Of special interest is the case of a crystalline undulator in which crystalline planes bending is periodic. In [1-3] the process of channeled electrons radiation in such a device was considered. It was shown that the radiation spectrum of particles in the crystal in this case contains maxima associated both with the period of oscillation of the channeled particles in the channel and with the period of the crystal planes bending. In [4] the radiation process in a crystalline undulator with large bending amplitude of the crystal planes was considered. Here the channeling phenomenon was absent. In this case the trajectories of particles that are crossing the crystal planes are close to rectilinear. This allowed consider the radiation process in the first approximation in the interaction of particles with the crystal field analytically. In that paper the possibility of manifesting coherent and interference effects in radiation leading to high radiation intensity was demonstrated. Taking into account the incoherent effects in the particle scattering leads to particle mutual transitions from the sub-barrier to the above-barrier states and vice versa. Analysis of the radiation process in this case can be carried out on the basis of numerical simulation method for this process. In this paper such method is developed on the basis of the proposed in [5] method of the radiation process simulation in a direct crystal. We present some results of simulation of 1 and 10 GeV electron radiation in crystalline undulator at different values of the plane bending amplitudes. The calculations are performed in the dipole approximation of radiation theory with taking into account the recoil effect in radiation.

KEYWORDS: Crystalline undulator; Coherent radiation; Numerical simulation of the radiation process.

---

[*] Corresponding author.

# Contents

## 1. Motion and coherent radiation of relativistic electrons in a crystalline undulator under conditions of above-barrier motion

The problem of the relativistic electron radiation in the field of periodically bent crystalline atomic planes (a crystalline undulator) was considered in [4] for the case when the plane bending amplitude was large ($x_0 \gg d_x$). In this case the particles will move mainly under conditions of above-barrier motion. On this condition the crystal field slightly perturbs the particle motion and its motion is close to rectilinear.

The motion of a relativistic electron at a small angle $\theta$ to the $z$ axis along which the atomic crystalline planes are deformed (see figure 1) is determined by the continuous potential of the curved crystalline plane:

$$U(x,z) = \sum_n u(x + x_0 Sin(\Omega z) - nd_x), \qquad (1)$$

where $u(x)$ is the continuous potential of a single non-curved crystalline plane, $d_x$ is the distance between the crystalline planes, $\Omega = 2\pi/T$, $T$ is the period of the plane bending along the $z$ axis and $x_0$ is the plane bending amplitude.

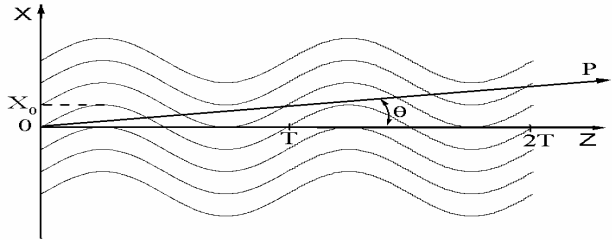

**Figure 1.** Motion of a relativistic electron at small θ angle to the ($y,z$) plane along which the atomic crystalline planes are bended.

In the case when the crystal field weakly perturbs the motion of the particle and its motion is close to rectilinear (this condition is satisfied if $x_0 \gg d_x$) the radiation cross section is determined by the Born formula:

$$d\sigma = \frac{e^2 \delta}{4\pi^3 m^2} \frac{d\omega}{\omega} \frac{\varepsilon'}{\varepsilon} d^3q \frac{q_\perp^2}{q_{II}^2} \left\{ 1 + \frac{\omega^2}{2\varepsilon\varepsilon'} - 2\frac{\delta}{q_{II}}\left(1 - \frac{\delta}{q_{II}}\right) \right\} |U_q|^2, \qquad (2)$$

where $\omega$ is the emitted photon energy, $m$ and $\varepsilon$ are the mass and energy of the initial electron, $\varepsilon' = \varepsilon - \omega$ is the energy of the final electron, $\delta = \omega m^2 / 2\varepsilon\varepsilon'$, $q_\perp$ and $q_{II}$ are the components of the recoil momentum to the external field and $U_q$ is the Fourier component of the potential of crystal planes.

In the problem under consideration the radiation cross section of a relativistic electron moving in a field (1) can be reduced to the form

$$\omega \frac{d\sigma}{d\omega} = L_x L_y L_z \frac{2e^2 \varepsilon' \delta}{m^2 a_x^2 \varepsilon} \sum_n \sum_{g_x} |U_{g_x}|^2 J_n^2(g_x x_0) \frac{g_x^2}{g_\parallel^2} \left\{ 1 + \frac{\omega^2}{2\varepsilon\varepsilon'} - 2\frac{\delta}{g_\parallel}\left(1 - \frac{\delta}{g_\parallel}\right) \right\}, \qquad (3)$$

where $L_x$, $L_y$ and $L_z$ are the linear dimensions of the crystal along the $x$, $y$ and $z$ axes, $J_n(g_x x_0)$ is the Bessel function of order $n$ and $g_x = 2\pi k/d_x$, $k$ is an integer. The summation in (3) is carried out over all values of $n$ and $k$ that satisfy condition $g_\parallel = \Omega n + \theta g_x \geq \delta$.

The radiation cross section (3) contains a large number of sharp maxima at frequencies satisfying the condition $\Omega n + \theta g_x = \delta$ (see figure 2).

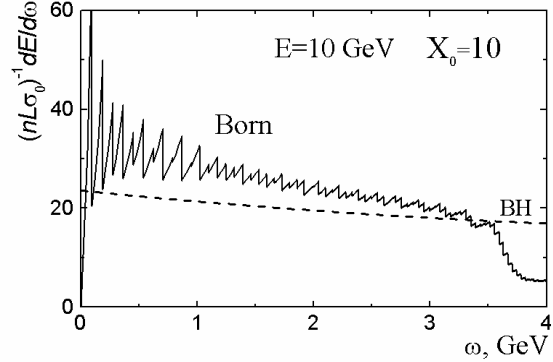

**Figure 2.** The radiation spectra of 10 GeV electron moving in the crystalline undulator field in the (110) silicon crystal plane at θ angle to the (y,z) plane equal zero. The plane bending amplitude is $X_0 = x_0/d_x = 10$, the bending period length is T=10 μm. The dotted curve is the Bethe-Heitler radiation spectrum.

## 2. Numerical simulation of motion and coherent radiation in a periodically bent crystal

In the case when the motion of a particle is not close to rectilinear and when the number of curves of the crystalline planes is finite the investigation can be carried out only by numerical simulation of the particle motion and radiation process in crystal.

The transverse component of a particle trajectory $x(t,x_0)$ at its movement in the crystal at small angle $\theta$ to the crystallographic plane is determined mainly in this case by the equation

$$\frac{d^2}{dt^2} x = -\frac{1}{\varepsilon} \frac{\partial}{\partial x} U(x,z), \qquad (4)$$

where $U(x,z)$ is the continuous potential of the crystal plane (1) and $z \approx vt$.

The solution of this equation is carried out on the basis of the method developed earlier at studying of ultrahigh-energy particles passage through a direct crystal [5].

The calculations of radiation spectrum carried out in the dipole approximation of quantum theory of radiation on the bases of the following formula:

$$\left\langle \frac{dE}{d\omega} \right\rangle = \frac{e^2 \omega}{4\pi} \int \frac{d^2 \boldsymbol{\rho}_0}{S} \int_\delta^\infty \frac{dq}{q^2} \left[ \frac{\varepsilon^2 + \varepsilon'^2}{\varepsilon\varepsilon'} - 4\frac{\delta}{q}\left(1 - \frac{\delta}{q}\right) \right] |W(q)|^2, \qquad (5)$$

where $\mathbf{W}(q) = \int_{-\infty}^{\infty} dt\, \dot{\mathbf{v}}_\perp(t) \exp(iqt)$, $\boldsymbol{\rho}_0$ is the transverse coordinate of the particle's entrance into the crystal and $S$ is the square of the area on which the beam falls.

To calculate the value of $\mathbf{W}(q)$, we used the previously developed program for numerical simulation of the passage of fast charged particle beams through a crystal near one of the crystalline axes [5]. According to this program, we divide the whole particle path in a crystal into a large number of intervals for which the collision's time $t_n$ and the scattering angle $\Delta\theta_n$ are determined by equation (4). So we have in this case $\mathbf{W}(q) = \sum_n e^{iqt_n}\Delta\boldsymbol{\theta}_n$.

The method of numerical simulation of radiation process in this case is developed in [5] with additional taking into account the incoherent effects in scattering. Due to this the particle transitions between the modes of their channeling and the above-barrier motion in the field of bent crystalline atom planes are possible.

The characteristics of relativistic electron radiation in a crystal strongly depend on the peculiarities of particle motion in a crystal. Some examples of trajectories for 1 GeV and 10 GeV electron moving in the field of crystalline undulator for different values of the plane bending amplitudes $X_0$ (it is assumed that the particles enter the silicon crystal at an angle $\theta = 0$ to the (110) plane) are presented in figure 3.

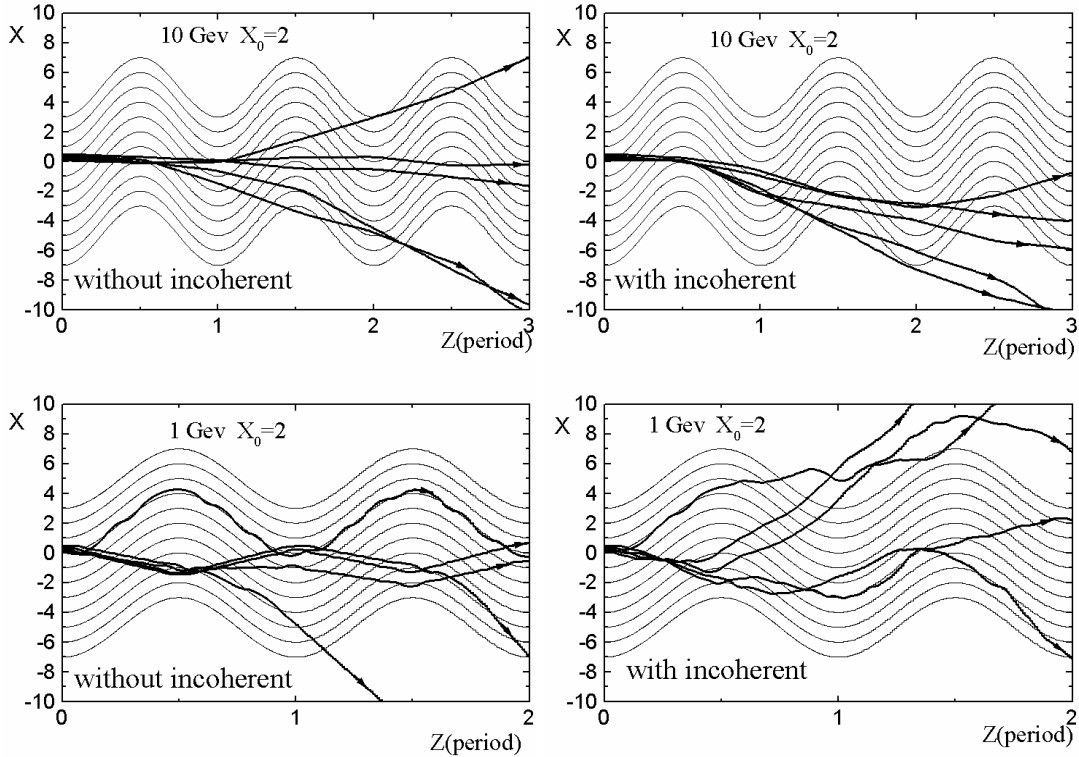

**Figure 3.** Typical trajectories of fast electrons in the field of crystalline undulator in the (x,z) plane. Different curves correspond to particles with different initial conditions of particles entry into the crystal, θ angle to the (y,z) plane equal zero. The plane bending period is T = 10 μm. The calculations are performed without (on the left) and with (on the right) taking into account incoherent scattering effects associated with the thermal vibration of atoms in the crystal lattice.

The above examples of fast electron trajectories in the field of crystalline undulator show that taking into account incoherent effects in the process of particle scattering by thermal vibrations of the crystal lattice atoms lead to significant changes in the trajectories even at the first bending periods of the undulator crystal planes.

The results of calculations of the spectral radiation density of 10 GeV electrons moving in the crystal undulator field in the (110) plane of the silicon crystal with the plane bending amplitude $X_0=10$ for crystal thickness equal to one plane bending period are presented in figure 4. In this case the radiation spectrum for particles moving in crystal without taking into account incoherent effects in the particle scattering by thermal vibrations of the atoms of the crystal lattice is in good agreement with the results of analytical calculations for above-barrier particles according to Born approximation formulas (3).

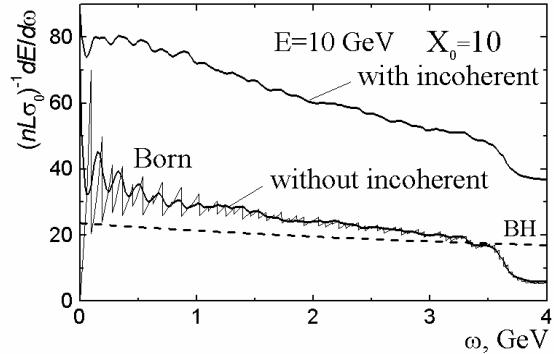

**Figure 4.** The radiation spectra of 10 GeV electrons moving in the crystalline undulator field in the (110) silicon crystal plane, θ angle to the (*y,z*) plane equal zero. The bending period length is T=10 μm. The thin curve is the result of the theory of coherent radiation, the thick curves are the results of the simulation, respectively, without and taking into account the incoherent particles scattering by thermal vibrations of the crystal lattice atoms, the dotted curve is the Bethe-Heitler radiation spectrum.

The fine structure of the radiation spectrum is somewhat different from the analytical calculations. This is due to the fact that analytical calculations refer to the case of a large number of bending of crystalline planes and numerical simulations are for one bending period of the plane. Taking into account incoherent effects in the particle scattering by thermal vibrations of the atoms of the crystal lattice leads to an additional contribution to the radiation which, however, in considered case is greater than the corresponding Bethe-Heitler result. This is due to the fact that in the case under consideration the beam enters the crystal along the crystalline planes (see figure 3).

Especially brightly the difference between numerical and analytical calculations (according to the Born approximation (3) formula) is manifested for small plane bending amplitudes and for smaller particle energies (see figure 5).

The results of numerical and analytical calculations of the spectral density of 10 and 1 GeV electron radiation moving in the crystal undulator field in the (110) plane of the silicon crystal for different plane bending amplitudes $X_0$ and for the crystal thickness equal to 5 plane bending periods are compared in figure 5. The calculations are performed with and without incoherent effects in scattering.

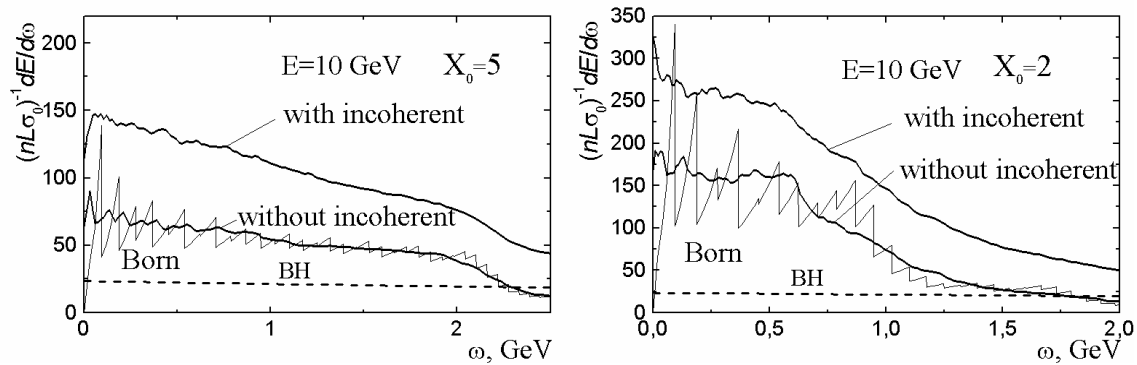

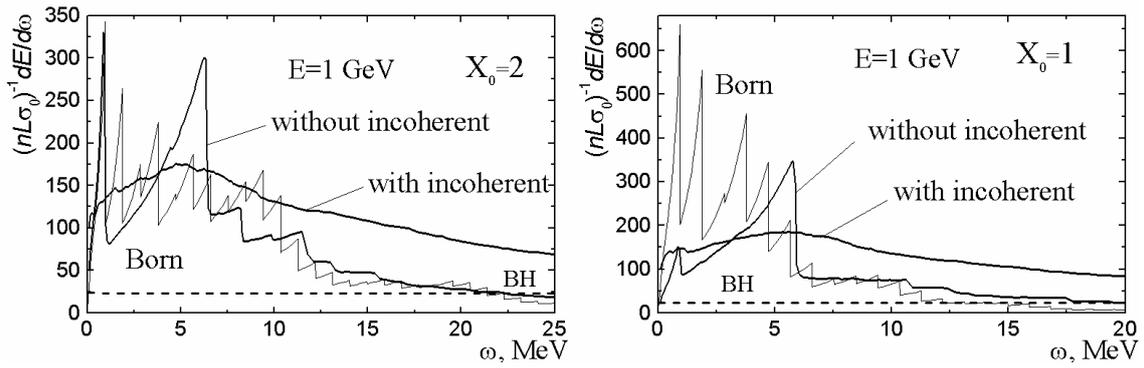

**Figure 5.** The radiation spectra of 10 and 1 GeV electrons moving in the crystalline undulator field in the (110) silicon crystal plane at θ angle to the (y,z) plane equal zero and at plane bending amplitudes $X_0 = 5$, 2, and 1. The bending period length is T=10 μm, the crystal thickness is 5 periods T. The thin curve is the result of the theory of coherent radiation [4], the thick curves are the results of the simulation, respectively, without and taking into account the incoherent particle scattering by thermal vibrations of the crystal lattice atoms, the dotted curve is the Bethe-Heitler radiation spectrum.

It is shown that the simulation results are in good agreement with the corresponding results of analytical calculations using the Born approximation (3) formulas for large values of plane bending amplitudes.

The obtained results demonstrate that taking into account incoherent effects in the process of particle scattering by thermal vibrations of the atoms of the crystal lattice leads to an additional contribution to the radiation. For small crystal thicknesses this contribution to the radiation can exceed the contribution of the corresponding Bethe-Heitler result.

**Acknowledgments**

The work is partly supported by the project of National Academy of Sciences of Ukraine No. CO-1-8.